\documentclass[12pt]{article}
\usepackage{amsfonts}
\usepackage{amsmath}
\usepackage{jeep}

\newtheorem{theorem}{Theorem}
\newtheorem{acknowledgement}[theorem]{Acknowledgement}

\input{tcilatex}
\begin{document}

\title{From Newton's Universal Gravitation to Einstein's Geometric Theory of
Gravity}
\author{Donald H. Kobe$^{1}$ and Ankit Srivastava$^{1,2}$ \\
%EndAName
$^{1}$Department of Physics\\
University of North Texas\\
Denton, Texas 76203\\
and\\
$^{2}$Department of Materials Science and Engineering\\
University of North Texas\\
Denton, Texas 76203}
\maketitle

\begin{abstract}
Starting with Newton's law of universal gravitation, we generalize it
step-by-step to obtain Einstein's geometric theory of gravity. Newton's
gravitational potential satisfies the Poisson equation. We relate the
potential to a component of the metric tensor by equating the
nonrelativistic result of the principle of stationary proper time to the
Lagrangian for a classical gravitational field. In the Poisson equation the
Laplacian of the component of the metric tensor is generalized to a cyclic
linear combination of the second derivatives of the metric tensor. In local
coordinates it is a single component of the Ricci tensor. This component of
the Ricci tensor is proportional to the mass density that is related to a
single component of the energy-momentum stress tensor and its trace. We thus
obtain a single component of Einstein's gravitational field equation in
local coordinates. From the principle of general covariance applied to a
single component, we obtain all tensor components of Einstein's
gravitational field equation.
\end{abstract}

\newpage

\section{Introduction}

\emph{\ \ }Issac Newton published his \textit{Principia} in 1686,\cite%
{principia} in which he formulated his three laws of motion and the law of
universal gravitation that together unified both celestial and terrestial
mechanics. He used these principles to derive Kepler's three laws of
planetary motion. In spite of the remarkable success of his mechanics, he
was not completely satisfied with his achievement. Newton himself criticized his
universal law of gravitation as it required the gravitational force to
act instantaneously at a distance without the presence of any medium to
transmit it.\cite{cohen} However, his gravitation was successfully applied
for over 200\emph{\ }years before it was superseded by Einstein.

In 1905 Einstein published his special theory of relativity \cite{einstein05}
that unified space and time and consequently made mechanics consistent with
electromagnetism. Einstein realized that Newton's principle of gravitation
was incompatible with the special theory of relativity. Newtonian mechanics presumes the
existence of an absolute space and an absolute time in which the propagation
of the gravitational interaction is instantaneous.\cite{principia} The
special theory of relativity postulates the constancy of the speed of light
that unifies the concepts of space and time into a four-dimensional
spacetime. Instantaneous propagation of the gravitational interaction in one
coordinate system would mean retarded or accelerated propagation in another.

Einstein began to work on the problem of gravity soon after he published the
special theory of relativity.\cite{norton} His first attempt was to modify
special relativity by using a speed of light depending on the gravitational
field, but it was unsuccessful. When he returned to Zurich from Prague in
1912 he began a new approach based on non-Euclidian geometry. His friend
Marcel Grossmann, a mathematician, helped him learn the \textquotedblleft
absolute differential calculus\textquotedblright\ (or tensor analysis as it
is now called) developed by Ricci and Levi-Civita.\cite{levi} Together they
formulated a new theory of gravitation. But later they rejected the first version for not having the correct Newtonian limit.\cite{pais} After much searching Einstein finally formulated a theory of gravity in 1915-1916
that was called the general theory of relativity.\cite{einstein16}

In most presentations of Einstein's general theory of relativity\cite{einstein50} his gravitational 
field equation is given in tensor form as
embodying some general principles. It is then shown that in the limit of
weak gravity and small speeds it reduces to Newtonian gravity.\cite{moller,
pauli, adler, bergmann} This top-down approach makes it difficult for the
student to understand the gravitational field equation. In this paper we use a 
bottom-up approach to obtain Einstein's field equation from Newton's universal gravitation. 
It should be noted that Einstein's field equation is not the only 
generalization of the Newton's universal gravitation but there could be other possible and more complicated 
generalizations as well. However we restrict ourself to obtain Einstein's field equation from Newton's universal gravitation.     

We begin with Poisson's equation and use a natural step-by-step
process of generalization. In the next section we review Newtonian gravity and Poisson's equation 
for the gravitational potential. In Sec. 3 Newton's gravitational potential is shown to be
related to the $00$-component of the metric tensor by equating the
nonrelativistic result of the principle of stationary proper time to the
classical Lagrangian for a particle in a gravitational potential. Poisson's
equation for the potential can then be written in terms of the second
derivative of the $00$-component of the metric. In Sec. 4 we
generalize this equation by using a cyclic linear combination of second
derivatives of the metric tensor to obtain the curvature of a
four-dimensional spacetime. The resultant is the $00$-component
of the Ricci tensor in local coordinates. In Sec. 5 the mass density in the Poisson
equation is related to the $00$-component of the energy-momentum stress
tensor and its trace.\cite{kobe82} Using the principle of general covariance \cite{einstein16} 
in Sec. 6, the single component of the tensor equation in local coordinates 
is generalized to obtain all components of the Einstein's gravitational field equation.
Our approach goes beyond other papers that make Einstein's gravitational field equation
plausible.\cite{chandra, krefetz, frankel} The conclusion is given in
Sec. 7 and the appendix gives a brief review of the Riemann and Ricci tensors.

\section{Poisson's Equation for Gravity}

Newton's universal law of gravity\cite{principia} states that the attractive
force $\mathbf{F}_{01}(\mathbf{x}_{0})$ on mass $m_{0}$ at $\mathbf{x}_{0}$
by a mass $m_{1}$ at $\mathbf{x}_{1}$ is given by 
\begin{equation}
\mathbf{F}_{01}\left( \mathbf{x}_{0}\right) =-Gm_{0}m_{1}\frac{\mathbf{\hat{x%
}}_{01}}{{|\mathbf{x}_{01}|}^{2}},  \label{F01}
\end{equation}%
where $G$ is Newton's universal gravitation constant, the displacement
between the particles is ${\mathbf{x}_{01}=\mathbf{x}}_{0}{-\mathbf{x}}_{1}$
and the corresponding unit vector is $\mathbf{\hat{x}}_{01}={\mathbf{x}%
_{01}/|\mathbf{x}_{01}|.}$

Because of the principle of superposition, the total gravitational force on
a test particle of mass $m_{0}$ at position $\mathbf{x}$ by particles of
mass $m_{i}$ at positions $\mathbf{x}_{i}$ ($i=1,2,\cdots ,N$) is

\begin{equation}
\mathbf{F}_{0}\left( \mathbf{x}\right) =-Gm_{0}\sum_{i=1}^{N}m_{i}\frac{%
\mathbf{x}-\mathbf{x}_{i}}{{|\mathbf{x}-\mathbf{x}}_{i}{|}^{3}}.
\label{super}
\end{equation}%
If the group of masses in Eq. (\ref{super}) is generalized to a continuous
distribution with a mass density $\rho (\mathbf{x^{\prime }})$ at the
position $\mathbf{x^{\prime }}$, then the gravitational field $\mathbf{g(x)}$
at the point $\mathbf{x,}$ defined as the force per unit test mass, is 
\begin{equation}
\mathbf{g}\left( \mathbf{x}\right) \equiv \frac{\mathbf{F}_{0}\left( \mathbf{%
x}\right) }{m_{0}}=-G\int d^{3}x^{\prime }\rho \left( \mathbf{x^{\prime }}%
\right) \frac{\mathbf{x}-\mathbf{x^{\prime }}}{{|\mathbf{x}-\mathbf{%
x^{\prime }}|}^{3}},  \label{super1}
\end{equation}%
where the volume integral on $\mathbf{x^{\prime }}$ is over all space. The
gravitational field $\mathbf{g}\left( \mathbf{x}\right) $ is regarded as the
carrier of the interaction.

From Eq. (\ref{super1}) the gravitational field is conservative since $%
\nabla \times \mathbf{g}\left( \mathbf{x}\right) =0$. Therefore a
gravitational potential function $\phi \left( \mathbf{x}\right) $ exists for
the gravitational field so that $\mathbf{g}\left( \mathbf{x}\right) =-\mathbf{\nabla 
}\phi \left( \mathbf{x}\right) $ and 
\begin{equation}
\phi \left( \mathbf{x}\right) =-G\int d^{3}x^{\prime }\frac{\rho \left( 
\mathbf{x^{\prime }}\right) }{|\mathbf{x}-\mathbf{x^{\prime }}|}.
\label{grav-pot}
\end{equation}%
Taking the Laplacian of Eq. (\ref{grav-pot}) gives Poisson's equation for
the Newtonian gravitational potential, 
\begin{equation}
\nabla ^{2}\phi \left( \mathbf{x}\right) =4\pi G\rho \left( \mathbf{x}%
\right) .  \label{poisson}
\end{equation}%
In a step-by-step process we will now generalize this equation to Einstein's
gravitational field equation. \ \ \ \ 

\section{Gravity and Geometry}

Einstein realized that since inertial mass and gravitational mass are equal
for gravity (the equivalence principle) it could be described by a
geometrical theory. In the general theory of relativity gravity is described
by the purely geometrical effect of motion in curved spacetime. The motion
of a free particle in the curved spacetime is determined from the principle
of stationary proper time\cite{ohanian} which states that along the
world line of a free particle the proper time is stationary. With our choice
of sign convention the stationary proper time is maximum for the true
trajectory compared to the proper time for all other trajectories.

Analogous to the Fermat's principle of least time in ray optics,\cite{feynman}
the principle of stationary proper time in relativity is%
\begin{equation}
\delta \tau =\delta \int_{\tau _{1}}^{\tau _{2}}d\tau =0,  \label{DT}
\end{equation}%
where $d\tau $ is the element of proper time in curved spacetime, the limits
of integration are arbitrary and the variation $\delta $ of the integral is
over all paths. The infinitesimal element of proper time $d\tau$ is 
\begin{equation}
d\tau =c^{-1}\left( g_{\mu \nu }dx^{\mu }dx^{\nu }\right) ^{1/2}>0,
\label{dtau}
\end{equation}%
where $\left(g_{\mu \nu }dx^{\mu }dx^{\nu }\right)^{1/2} = ds$ is an element of spacetime separation and $c$ is the speed of light. In a curved spacetime the metric tensor is $g_{\mu \nu }=g_{\mu \nu
}\left( x\right) ,$ where the spacetime coordinates are $x=x^{\mu }=\left(
x^{0},x^{1},x^{2},x^{3}\right) ,$ the time coordinate is $x^{0}=ct$ and the
spatial coordinates are $\mathbf{x=(}x^{1}$, $x^{2}$, $x^{3})$. In special
relativity the metric tensor is diagonal and we choose the convention $%
g_{\mu \nu }=\eta _{\mu \nu }=diag\left\{ 1,-1,-1,-1\right\} .$

Substituting the element of proper time, Eq. (\ref{dtau}), into Eq. (\ref%
{DT}) and, multiplying and dividing it by $dt$ gives 
\begin{equation}
\delta \tau =c^{-1}\delta \int \left( g_{\mu \nu }\frac{dx^{\mu }}{dt}\frac{%
dx^{\nu }}{dt}\right) ^{1/2}dt=0.  \label{sta-1}
\end{equation}%
In a weak gravitational field we can choose a local coordinate system for
which the metric is $g_{ij}=-\delta _{ij},$ $g_{0i}=0$ and $g_{00}=g_{00}\left( 
\mathbf{x}\right) ,$ where $\delta _{ij}$ is the Kronecker delta, repeated
Greek indices are summed from $0$ to $3,$ and repeated Latin indices are
summed over spatial values from $1$ to $3$. With this choice of coordinates,
Eq. (\ref{sta-1}) can be rewritten as 
\begin{equation}
\delta \tau =c^{-1}\delta \int \left( g_{00}\frac{dx^{0}}{dt}\frac{dx^{0}}{dt%
}-\delta _{ij}\frac{dx^{i}}{dt}\frac{dx^{j}}{dt}\right) ^{1/2}dt=0.
\label{sta-2}
\end{equation}%
This equation can be simplified by using $dx^{0}/dt=c$ so that Eq. (\ref{sta-2})
becomes, 
\begin{equation}
\delta \tau =\delta \int \left( g_{00}-\frac{\mathbf{\dot{x}}^{2}}{c^{2}}%
\right) ^{1/2}dt=0.  \label{sta3}
\end{equation}%
We assume speed to be small compared to the speed of light $|\mathbf{\dot{x}|<<}%
c $ and a weak gravitational field so that $g_{00}=1+h_{00},$ where $%
h_{00}<<1. $ Then a Taylor series expansion of the square root in Eq. (\ref%
{sta3}) to first order gives 
\begin{equation}
\delta \tau =\delta \int \frac{1}{2}\left( h_{00}-\frac{\mathbf{\dot{x}}^{2}%
}{c^{2}}\right) dt=0.  \label{sta5}
\end{equation}%
Multiplying Eq. (\ref{sta5}) by $-mc^{2}$ leads to the
principle of least action 
\begin{equation}
\delta S=-mc^{2}\delta \tau =\delta \int \left( \frac{1}{2}m\mathbf{\dot{x}}%
^{2}-\frac{1}{2}mc^{2}h_{00}\right) dt=0,  \label{sta6}
\end{equation}%
where the integrand can be interpreted as the Lagrangian for a particle in a weak
gravitational field. Equating this Lagrangian with the classical Lagrangian $%
L_{\phi }$ for a particle in a gravitational potential $\phi ,$%
\begin{equation}
L_{\phi }=\frac{1}{2}m\mathbf{\dot{x}}^{2}-m\phi \left( \mathbf{x}\right) ,
\label{Lphi}
\end{equation}%
we obtain a relationship between $h_{00}$ and $\phi ,$ 
\begin{equation}
h_{00}=g_{00}-1=\frac{2\phi }{c^{2}}.  \label{connection}
\end{equation}%
We have now established a relationship between the component $g_{00}$ of the
metric tensor and the gravitational potential $\phi $. The relation in Eq. (%
\ref{connection}) has been previously derived,\cite{kobe82, LL} but by
different methods. In most of the text books this relation is obtained 
from the Einstein's gravitational field equation in the Newtonian limit.

\section{Gravitational Curvature}

We have shown in Eq. (\ref{connection}) that the gravitational potential $%
\phi $ can be replaced by the $00$-component of the metric under
nonrelativistic assumptions. Substituting this relation into Poisson's
equation (\ref{poisson}), we get 
\begin{equation}
g_{00,i,i}=-g_{00,i}^{~~~~~,i}=\frac{8\pi G}{c^{2}}\rho   \label{poi-con}
\end{equation}%
here we use the comma notation for partial derivatives and
Einstein's summation convention over repeated Latin indices which varies from $1$ to $3$.
For later generalization to tensor contraction we distinguish between
covariant and contravariant indices. To obtain a contravariant index we
raise an index using $g_{00,i}^{~~~~~,i}=g_{00,i,j}g^{ji}=-g_{00,i,i}$,
where $g^{ji}=-\delta ^{ji}$ in our convention for the metric in special
relativity. 

The Laplacian of $g_{00}$ measures the intrinsic curvature of
three-dimensional space. The intrinsic curvature of four-dimensional spacetime must involve the
second derivatives of the metric component $g_{00}$ with respect to all
spacetime coordinates.\cite{hakim, love} The generalization first requires: 
changing the double Latin indices $i$ for derivatives in Eq. (%
\ref{poi-con}) to double Greek indices $\mu $ that are varies from $0$ to $3$%
,%
\begin{equation}
g_{00,i}^{~~~~~,i}\Longrightarrow g_{00,\mu }^{~~~~~~,\mu }=g^{\mu \nu
}g_{00,\mu ,\nu }\text{ .}  \label{g00uv}
\end{equation}%

Since $g_{00}(x)$ depends on all the spacetime coordinates, we now have to
include a linear combination of all its second derivatives to find the
intrinsic curvature. This further generalization requires, 
finding the appropriate linear combination of second derivatives of the
metric tensor. Also the terms in the linear combination should not favor any
particular one, so that each term has the same amplitude. The terms in the linear 
combination can have different signs, but they should have equal number of positive and
negative terms. A natural choice for the right-hand side of Eq. (\ref{g00uv}%
) that satisfies these requirements is to replace the indices $00\mu \nu $\
by the sum over their cyclic permutations $\left( 00\mu \nu \right) $\ with
the appropriate sign of the permutation, 
\begin{equation}
g^{\mu \nu }g_{00,\mu ,\nu }\Longrightarrow g^{\mu \nu
}\sum_{cyclic}\epsilon g_{00,\mu ,\nu }\text{ .}  \label{cyclic}
\end{equation}%
The summation on the right-hand side is defined\cite{bostock} as the sum of
cyclic permutations of $00\mu \nu $ with the appropriate sign $\epsilon
=+(-) $ for an even (odd) permutation, so Eq. (\ref{cyclic}) is%
\begin{equation}
g^{\mu \nu }\left( g_{00,\mu ,\nu }-g_{0\mu ,\nu ,0}+g_{\mu \nu ,0,0}-g_{\nu
0,0,\mu }\right) =2R_{00,}  \label{R00}
\end{equation}%
in the convention for the Ricci tensor of Rindler\cite{rindler} and Ohanian.%
\cite{ohanian} For static Newtonian gravity all the terms on the left-hand
side of Eq. (\ref{R00}) that involve a time derivative vanish. The terms on
the left-hand side of Eq. (\ref{R00}) are equal to twice $R_{00},$ the $00$%
-component of the\ Ricci tensor in local coordinates.\cite{ohanian, rindler}%
\emph{\ }Using the generalizations in Eqs. (\ref{g00uv}), (\ref{cyclic}) and
(\ref{R00}) in the left-hand side of the Poisson equation (\ref{poi-con}),
we obtain%
\begin{equation}
R_{00}=-\frac{4\pi G}{c^{2}}\rho .  \label{poisson2}
\end{equation}

One of the subscripts of the $00$-component of the Ricci tensor in local
coordinates can be raised with a contravariant metric tensor, so $%
R_{0}^{~0}=g^{0\alpha }R_{0\alpha },$ where $g^{0\alpha }=\delta ^{0\alpha
}. $ Thus raising a $0$-index does not change the numerical value of the
tensor component in local coordinates. Hence, Eq. (\ref{poisson2}) now
becomes 
\begin{equation}
R_{0}^{~0}=-\frac{4\pi G}{c^{4}}\rho c^{2}.  \label{poi-ricci}
\end{equation}%
Since the left-hand side of Eq. (\ref{poi-ricci}) is a component of a
tensor, the right-hand side of the equation must also be the same component
of another tensor.\bigskip

\section{Energy-Momentum Stress Tensor}

On the right-hand side of Eq. (\ref{poi-ricci}) the term $\rho c^{2}$ is the
mass-energy density. In the cloud of dust model\cite{ohanian} it is related
to a component of the stress tensor. The energy-momentum stress tensor $%
T_{\mu }^{~~\nu }$ for a static tenuous gas of density $\rho $ and
negligible pressure has a component $T_{0}^{~0}$ that is the rest energy
density $\rho c^{2}$.\cite{adler, kobe82} The other components of $T_{\mu
}^{~~\nu }$ are negligible in this model, so the trace $T=T_{\mu }^{~\mu }$
of the stress tensor is also $\rho c^{2}$. Since the trace $T$ is a scalar,
hence the simplest second-rank tensor that can be constructed from it is $T\delta
_{\mu }^{~\nu }$, where $\delta _{\mu }^{~\nu }$ is the Kronecker delta
tensor.

Hence the Newtonian mass density $\rho $ times $c^{2}$ can be generalized to
a linear combination of $T_{0}^{~0}$ and $T\delta _{0}^{~0}$ for all matter
and fields, 
\begin{equation}
\rho c^{2}=aT_{0}^{~0}+\left( 1-a\right) T\delta _{0}^{~0}  \label{lin-com}
\end{equation}%
where $a$ is a constant to be determined later. Substituting Eq. (\ref%
{lin-com}) in Eq. (\ref{poi-ricci}), we get 
\begin{equation}
R_{0}^{~0}=-\frac{4\pi G}{c^{4}}\left( aT_{0}^{~0}+\left( 1-a\right) T\delta
_{0}^{~0}\right) .  \label{EFE1}
\end{equation}%
Since both sides of this equation are $00$-components of a tensor equation,
we can generalize it to obtain all the other components.

\section{Einstein's Gravitational Field Equation}

Equation (\ref{EFE1}) has only one component of the Ricci and stress-energy
tensors. Nevertheless, it establishes a connection between the gravitational
field and the geometry of spacetime under nonrelativistic assumptions in a
local coordinate system. Using the principle of general covariance, we can
generalize Eq. (\ref{EFE1}) to obtain the complete Einstein gravitational
field equation. Einstein struggled between 1912-1916 to find the correct formulation of the
covariance principle.\cite{earman} The principle of general covariance as formulated by Einstein%
\cite{einstein16} states that, \textit{\textquotedblleft The general laws of
nature are to be expressed by equations which hold good for all systems of
coordinates, ....\textquotedblright } This principle means that laws of
physics must be tensor equations, \textit{i.e.}, equations that do not
change their form under general coordinate transformations. 

In Sec. 3 the metric tensor is assumed to be static with the off-diagonal
terms zero. Under a general coordinate transformation the metric tensor can
have all components $g_{\mu \nu }(x)$ and be a function of all the spacetime
coordinates $x=x^{\mu }$. Applying the principle of general covariance to
Eq. (\ref{EFE1}), we get the full tensor Einstein gravitational field
equation valid in all coordinate systems, 
\begin{equation}
R_{\mu }^{~\nu }=-\frac{4\pi G}{c^{4}}\left( aT_{\mu }^{~\nu }+\left(
1-a\right) T\delta _{\mu }^{~\nu }\right) ,  \label{EFE2}
\end{equation}%
except for an undetermined constant $a$. The Ricci tensor $R_{\mu }^{~\nu }$
in Eq. (\ref{EFE2}) in its general form involves both the first and second
derivatives of the metric tensor. The energy-momentum stress tensor $T_{\mu
}^{~\nu }$ is generalized to include all material and non-gravitational
fields.\emph{\ }

We can now determine the constant $a$ in Eq. (\ref{EFE2}). The trace of Eq. (%
\ref{EFE2}) gives 
\begin{equation}
R=-\frac{4\pi G}{c^{4}}T\left( 4-3a\right) ,  \label{trace}
\end{equation}%
where the trace $\delta _{\mu }^{~\mu }=4$ and the scalar curvature is $%
R_{\mu }^{~\mu }=R$. Hence we can rewrite Eq. (\ref{EFE2}) in terms of $R$
instead of $T$,

\begin{equation}
R_{\mu }^{~\nu }-\frac{\left( 1-a\right) }{\left( 4-3a\right) }\delta _{\mu
}^{~\nu }R=-\frac{4a\pi G}{c^{4}}T_{\mu }^{~\nu }.  \label{EFE4}
\end{equation}

To determine $a$ we can take the covariant derivative of Eq. (\ref{EFE4}).
Because of \ the law of conservation for energy-momentum\cite{MTW} the
covariant derivative of the right-hand side of Eq. (\ref{EFE4}) vanishes 
\begin{equation}
T_{\mu ~;\nu }^{~\nu }=0,  \label{rhs}
\end{equation}%
where the semicolon denotes the covariant derivative. However, the covariant
derivative reduces to the partial derivative in local coordinates. The
covariant derivative of the left-hand side will also vanish if we use the
second Bianchi identity, 
\begin{equation}
\left( R_{\mu }^{~\nu }-\frac{1}{2}\delta _{\mu }^{~\nu }R\right) _{;\nu }=0,
\label{2B}
\end{equation}
Since the covariant derivative of the right-hand side of Eq. (\ref{EFE4})
vanishes, the coefficient of $\delta _{\mu }^{~\nu }R$ must be equal to $1/2$
which gives $a=2$.

Hence substituting the value of $a$ in Eq. (\ref{EFE4}) leads to the Einstein's
gravitational field equation\cite{einstein16} 
\begin{equation}
R_{\mu }^{~\nu }-\frac{1}{2}\delta _{\mu }^{~\nu }R=-\frac{8\pi G}{c^{4}}%
T_{\mu }^{~\nu }\text{ \ .}  \label{EFE}
\end{equation}%

The left-hand side of Eq. (\ref{EFE}) is defined as the \textit{Einstein
tensor} $G_{\mu }^{~\nu }$, 
\begin{equation}
G_{\mu }^{~\nu }\equiv R_{\mu }^{~\nu }-\frac{1}{2}\delta _{\mu }^{~\nu }R.
\label{G}
\end{equation}%
Hence the Einstein's gravitational field equation (\ref{EFE}) can be
rewritten in a more compact form as 
\begin{equation}
G_{\mu \nu }=-\frac{8\pi G}{c^{4}}T_{\mu \nu }\text{ .}  \label{EFE5}
\end{equation}%
The tensor $G_{\mu \nu }$ is the only tensor that can be constructed from $%
g_{\mu \nu }$, and its first and second partial derivatives whose
four-divergence vanishes.\cite{lovelock}

Einstein's equation was very successful in predicting the bending of light
by the sun, the precession of the perihelion of Mercury and radar echo delay.%
\cite{rindler, weinberg} Another prediction of Einstein's equation is
gravitation waves.\cite{MTW, schutz} Although they have not yet been
directly observed, but have been inferred from the loss of energy in binary
pulsar systems.\cite{hulse}

When Einstein first applied his field equation to cosmology in 1917 he found
that his equation did not predicted a static universe that was expected at
the time. He then introduced another term $\Lambda $, which he called the 
\textit{cosmological constant}, into his equation,\cite{einstein17}%
\begin{equation}
R_{\mu }^{~\nu }-\frac{1}{2}\delta _{\mu }^{~\nu }R-\Lambda \delta _{\mu
}^{~\nu }=-\frac{8\pi G}{c^{4}}T_{\mu }^{~\nu }\text{ \ ,}  \label{Lambda}
\end{equation}%
but the resulting static universe was not stable. When Hubble in 1929
established that the universe was expanding,\cite{hubble} Einstein called
the cosmological constant \textquotedblleft the biggest blunder of his
life.\textquotedblright\ If he had stuck with his original equation in 1917
he could have predicted the expansion of the universe over a decade before
it was discovered. Today the cosmological constant has reemerged as one
explanation for dark energy.\cite{peebles}

\section{Conclusion}

In this paper Einstein's gravitational field equation is obtained from a
step-by-step generalization of Newtonian gravitation. We first show that
Newton's law of universal gravitation leads directly to Poisson's equation
that relates the Laplacian of the gravitational potential to the mass
density. Then we use the principle of stationary proper time under
nonrelativistic assumptions and compare it with the classical Lagrangian of
a particle in a gravitational potential. We find a linear relationship
between the $00$-component of the metric tensor and the Newtonian
gravitational potential.

We then replace the Laplacian of the gravitational potential in Poisson's
equation (\ref{poisson}) with the Laplacian of the $00$-component of the
metric tensor $g_{00,i,i}.$ The Latin indices in three-dimensions are
generalized to Greek indices in four-dimensions $g^{\mu \nu }g_{00,\mu ,\nu
} $, where $g^{\mu \nu }$ is used to raise one of the indices to give a
tensor contraction. The indices of $00\mu \nu $ are generalized to give a
cyclic linear combination $\left( 00\mu \nu \right) $ for the indices. When
it is multiplied by $g^{\mu \nu }$ to raise an index, the linear combination
results in twice the $00$-component of the Ricci tensor in local coordinates.

The Poisson equation shows that the $00$-component of the Ricci tensor (\ref%
{R00}) is proportional to the mass density $\rho .$ Using the cloud of dust
model, we show that the mass-energy density $\rho c^{2}$ is proportional to
both the $00$-component of the energy-momentum stress tensor and the trace
of the tensor. Hence the $00$-component of the Ricci tensor is proportional
to a linear combination of the $00$-component of the energy-momentum stress
tensor and the $00$-component of its trace times the unit tensor. Einstein's
principle of general covariance states that the equations of physics must be
tensor equations, so that equations obtained in local coordinate systems for
one component must also be valid in all coordinate systems for all
components. Therefore, the full Ricci tensor is proportional to a linear
combination of the full energy-momentum stress tensor and its trace times the
unit tensor. The constant in the linear combination is obtained by using
energy-momentum conservation and the second Bianchi identity. The result is
Einstein's equation for the gravitational field.

Einstein's theory of gravitation has the reputation that it 
is a subject reserved only for graduate school. Wheeler\cite{wheeler} has
said that there is a need for a \textquotedblleft
quasi-Newtonian\textquotedblright\ argument for the correct field equations.
A previous paper\cite{kobe82} was an attempt to give such a bottom-up
derivation, but there were some gaps in the argument. This paper fills those
gaps. The approach used in this paper should make general relativity
accessible to undergraduate students. The subject can be taught at several
different levels. For a survey of the subject a sketch of the approach can
be used. At an intermediate level more details can be included. At an
advanced level all the mathematics can be included. We hope that this
approach will become a standard method for teaching general relativity and
giving all students a better understanding of Einstein's theory of
gravitation.

\textit{\ }

\begin{acknowledgement}
D.H.K. acknowledges discussions with Michael E. Burns. A.S. thanks Dr. A.
Needleman, Professor, University of North Texas, for his support while
carrying out this work.
\end{acknowledgement}

\section{Appendix: Riemann and Ricci Tensors}

The most important tensor in Riemann geometry is the Riemann curvature
tensor. The fully covariant Riemann curvature tensor in four-dimensional
curved spacetime is,\cite{ohanian, rindler} 
\begin{equation}
R_{\mu \alpha \nu \beta }=-\Gamma _{\mu \alpha \nu ,\beta }+\Gamma _{\mu
\alpha \beta ,\nu }+\Gamma _{~\alpha \nu }^{\gamma }\Gamma _{\gamma \mu
\beta }-\Gamma _{~\alpha \beta }^{\gamma }\Gamma _{\gamma \mu \nu \text{ \ }%
},  \label{rei}
\end{equation}%
where the Christoffel symbols of the first kind are%
\begin{equation}
\Gamma _{\alpha \beta \gamma }=\frac{1}{2}\left( g_{\alpha \beta ,\gamma
}-g_{\beta \gamma ,\alpha }+g_{\gamma \alpha ,\beta }\right) ,  \label{CS1}
\end{equation}%
and the second kind are 
\begin{equation}
\Gamma _{~\beta \gamma }^{\alpha }=g^{\alpha \delta }\Gamma _{\delta \beta
\gamma }=\frac{1}{2}g^{\alpha \delta }\left( g_{\delta \beta ,\gamma
}-g_{\beta \gamma ,\delta }+g_{\gamma \delta ,\beta }\right) \text{ \ .}
\label{CS2}
\end{equation}

We use local coordinates in which the first derivatives of the metric tensor 
$g_{\alpha \beta ,\gamma }$ vanish, so the Riemann tensor reduces to%
\begin{eqnarray}
R_{\mu \alpha \nu \beta } &=&-\Gamma _{\mu \alpha \nu ,\beta }+\Gamma _{\mu
\alpha \beta ,\nu }  \notag \\
R_{\mu \alpha \nu \beta } &=&\text{ }-\frac{1}{2}\left( g_{\mu \alpha ,\nu
,\beta }-g_{\alpha \nu ,\mu ,\beta }+g_{\nu \mu ,\alpha ,\beta }\right) \,+%
\frac{1}{2}\left( g_{\mu \alpha ,\beta ,\nu }-g_{\alpha \beta ,\mu ,\nu
}+g_{\beta \mu ,\alpha ,\nu }\right) .  \label{rie-local}
\end{eqnarray}%
The two partial derivatives commute, the metric tensor is symmetric $%
g_{\beta \mu }=g_{\mu \beta }$ and $R_{\mu \alpha \nu \beta }=-R_{\mu \alpha
\beta \nu }$, so this equation reduces to%
\begin{equation}
R_{\mu \alpha \beta \nu }=-\frac{1}{2}\left( g_{\mu \beta ,\alpha ,\nu
}-g_{\beta \alpha ,\nu ,\mu }+g_{\alpha \nu ,\mu ,\beta }-g_{\nu \mu ,\beta
,\alpha }\right) .  \label{rie-local2}
\end{equation}%
The indices on the right-hand side are cyclic $\left( \mu \beta \alpha \nu
\right) .$

The Ricci tensor is defined as\cite{ohanian, rindler}%
\begin{equation}
R_{\alpha \beta }\equiv R_{~~\alpha \beta \mu }^{\mu }.  \label{ricci}
\end{equation}%
Therefore, in local coordinates we get

\begin{eqnarray}
g^{\mu \nu }R_{\mu \alpha \beta \nu } &=&\text{ }-\frac{1}{2}g^{\mu \nu
}\left( g_{\mu \beta ,\alpha ,\nu }-g_{\beta \alpha ,\nu ,\mu }+g_{\alpha
\nu ,\mu ,\beta }-g_{\nu \mu ,\beta ,\alpha }\right) ,  \notag \\
R_{\alpha \beta } &=&\frac{1}{2}g^{\mu \nu }\left( g_{\beta \alpha ,\nu ,\mu
}-g_{\alpha \nu ,\mu ,\beta }+g_{\nu \mu ,\beta ,\alpha }-g_{\mu \beta
,\alpha ,\nu }\right) ,  \label{riccilocal}
\end{eqnarray}%
where the indices of the terms in parentheses in the last line are cyclic.
In Eq. (\ref{R00}) we use the $00$-component of the Ricci tensor in local
coordinates. By inspection the Ricci tensor (\ref{riccilocal}) in local
coordinates is symmetric,%
\begin{equation}
R_{\alpha \beta }=R_{\alpha \beta },  \label{sym}
\end{equation}%
and by the principle of general covariance this property holds in general
coordinates.

$^{a)}$Electronic mail: kobe@unt.edu

$^{b)}$Electronic mail: AnkitSrivastava@my.unt.edu

\end{document}